\documentclass[11pt,twoside]{article}
\usepackage{./asp2014}

\aspSuppressVolSlug
\resetcounters

\begin{document}

\title{The LSST Data Processing Software Stack: Summer 2015 Release}
\author{Tim~Jenness,$^1$ for the LSST Data Management Team
\affil{$^1$Large Synoptic Survey Telescope, Tucson, AZ, USA; \email{tjenness@lsst.org}}}

\paperauthor{Tim~Jenness}{tjenness@lsst.org}{0000-0001-5982-167X}{LSST}{Data Management}{Tucson}{AZ}{85719}{USA}

\begin{abstract}
  The Large Synoptic Survey Telescope (LSST) is an 8-m optical
  ground-based telescope being constructed on Cerro Pach\'on in
  Chile. LSST will survey half the sky every few nights in six optical
  bands. The data will be transferred to NCSA and within 60 seconds
  they will be reduced using difference imaging techniques and
  detected transients will be announced to the community in the
  VOEvent format. Annual data releases will be made from all the data
  during the 10-year mission, with unprecedented depth of coadds and
  time resolution of catalogs for such a large region of sky. In this
  paper we present the current status of the data processing software,
  and describe how to obtain it.
\end{abstract}

\section{The Science Pipelines Software ``stack''}

The Large Synoptic Survey Telescope
\citep[LSST;][]{2008arXiv0805.2366I} will take about 15\,TB of image
data per night and after ten years of operations will have
15\,petabytes of catalog data for the final data release and
0.5\,exabytes of image
data\footnote{\url{http://lsst.org/scientists/keynumbers}}. We are
writing a suite of software packages to enable these data products to
be created with sufficient quality and performance to meet the
established science goals \citep{2009arXiv0912.0201L}.

The science pipeline software will enable two key components of the data
management system. The Alert Production pipelines (also known as
\emph{Level 1}) process the data from the telescope and publish alerts
to the community within 60 seconds of data acquisition \citep{2014htu..conf...19K}. Data Release
Production (\emph{Level 2}) is responsible for the annual data
releases which reprocess all the data each year to generate the best
possible catalogs. Both these systems will be integrated with the
Calibration Products Production that continuously calculates the best
calibrations for the pipelines. The software will also provide a toolkit
for user-supplied code that can be used to efficiently and effectively
analyze LSST data as part of \emph{Level 3} processing or their own
pipelines. Full details of the data management applications design are
detailed elsewhere \citep{O3-1_adassxxv,LDM-151}.

The LSST data management science pipelines software system, commonly
referred to as the ``stack'', is a collection of about 40 separate
packages providing functionality such as data access libraries, data
models representing exposures and catalogs, source detection
algorithms, astrometry fitting, and photometry and measurement
algorithms. The software is written in a mixture of Python and
C++\footnote{Currently we support Python 2.7 but intend to also
  support Python 3. We are also migrating the C++ codebase to C++11.},
where the latter is used for CPU-intensive algorithms, or when the
algorithms require access to complex data structures. The codebase
consists of approximately 100,000 lines of Python and 110,000 lines of
C++ (not including comment or blank lines and not counting expanded
SWIG \citep[see e.g.][]{beazley2003automated} interface code).

The science pipeline packages are namespaced (with an \texttt{lsst.}\
root) and grouped by their functionality. The core namespaces are
defined as follows:

\begin{description}
\item[\textbf{daf}] The Data Access Framework is responsible for
  mediating between the archive resources and the application
  writer. The pipeline code has a completely abstract view of
  file I/O and only has to know how to deal with data objects
  representing fundamental types such as exposures and
  tables. Currently FITS is the internal format but the system is
  designed such that the internal format could be changed to HDF5, for
  example, and no changes would have to be made to the science
  pipeline code. This abstraction of the files from the code protects
  us against shifts in community preferences such as those discussed
  in \citet{2015ASPC..495...11M}.

\item[\textbf{afw}] The Astronomy FrameWork provides the
  core classes for manipulating exposures and catalogs, including
  detecting sources and world coordinate handling.

\item[\textbf{ip}] These are the image processing classes, including
  packages for instrument signature removal and image differencing.

\item[\textbf{meas}] The measurement packages include code for
  determining source properties and correcting astrometry and photometry.

\item[\textbf{obs}] These classes provide instrument-specific
  knowledge to the software system, providing information to the data
  access framework to teach it how to interpret data from a range of
  optical cameras. The \texttt{obs} packages currently support data
  from some instruments on Subaru and CFHT, in addition to simulated
  LSST data. Work is ongoing to add support for DECam.

\item[\textbf{pipe}] Pipeline infrastructure and tasks. A task is the
  name for a core processing component that can be chained with other
  tasks to build a pipeline.

\end{description}

More details concerning the history behind the development of the
LSST software can be found in \citet{2010SPIE.7740E..15A}.

\section{Summer 2015 release}

Whilst the software is open
source\footnote{\url{https://github.com/lsst}} and can be installed at
any time, LSST makes formal releases of the science pipeline software
at the end of each six month development cycle in the spring and
autumn. The most recent release covered the summer development cycle
and was labeled \emph{Summer 2015} and released in September
2015. Detailed release notes can be found
online;\footnote{\url{https://community.lsst.org/t/268}} here we
provide a summary.

\paragraph{Multi-band processing for coadds} New command-line tasks
have been added for consistent multi-band processing of coadds. This
new data processing flow carefully combines source measurements taken
in multiple bands to guarantee consistent deblending across all bands,
including when carrying out forced photometry, thereby enabling
reliable color measurement, and ensuring that all sources are
measured in each band, regardless of the bands where they are detected.

\paragraph{Upgraded astrometry calculation} Previously astrometry was
calculated using astrometry.net code \citep[][ascl:1208.001]{2010AJ....139.1782L} and
related catalogs distributed by LSST. To improve flexibility in the
code the astrometry fitter is now pluggable and now includes an
alternative implementation.

\paragraph{Support for PSFEx} PSFEx (ascl:1301.001) is currently the state of the art
external package for point spread function (PSF) determination, used
in projects such as DES \citep{2011ASPC..442..435B}. LSST wrappers
were created such that PSFEx could be used as a plugin in place of the
built in PSF determiner.

\paragraph{More efficient handling of large footprints} A footprint
defines the pixels associated with a particular source or blended
sources. This release saw significant improvements in performance when
using very large footprints.

\paragraph{Enable use of deblended heavy footprints in coadd forced
  photometry} Given the new multi-band processing for coadds we now
have a reference catalog that is consistent across all bands. This
catalog allows the use of the source's heavy footprints\footnote{A
  heavy footprint is a footprint that includes the pixel values.} to
replace neighbors with noise in forced photometry, thus providing
deblended forced photometry and consistent deblending across all
bands. This provides much better colors for blended objects as well as
measurements for drop-out objects that do not get detected in the
canonical band. This functionality has been enabled for forced coadd
photometry.

\paragraph{Significant improvements in the table class} The AFW package
has a native C++ implementation of a class for manipulating table data
for handling the results of detection and measurement
algorithms. This release comes with some major enhancements to the
internals of \texttt{afw.table} and, in particular, much better
support for compound fields (such as Right Ascension/Declination tuples).

\paragraph{Device independent displays} DS9
\citep[][ascl:0003.002]{2011ASPC..442..633J} is no longer hard-wired into the software
and the choice of display tool is now user configurable.  The
intention is for the next release to include support for the Firefly
visualization tool \citep{O10-1_adassxxv}.

\section{Obtaining the software}

The software is known to work on CentOS 6 and 7, recent Debians and
Mac OS X Yosemite and Mavericks (this release is known not to
  work on Mac OS X El Capitan due to interactions between library path
  environment variables and the new System Integrity Protection
  feature; this has been fixed in the current development version), with a C++11 compatible compiler such as GCC 4.8.3 or later, or
Apple \texttt{clang} 6 . The recommended way to install the software from source is to
use the \texttt{eups} distribution installation system
\citep{EUPS}. Experimental binary releases have also been
made available using a CernVM File
System \citep[CernVM-FS;][]{2015JPhCS.608a2031M}. Full details on both
these options are available on the release notes page.

\acknowledgements The Summer 2015 release of the LSST software stack
is the result of the efforts of the many people who are part of the
Data Management Team at LSST, as well as outside contributors.  This
material is based upon work supported in part by the National Science
Foundation through Cooperative Support Agreement (CSA) Award
No. AST-1227061 under Governing Cooperative Agreement 1258333 managed
by the Association of Universities for Research in Astronomy (AURA),
and the Department of Energy under Contract No. DE-AC02-76SF00515 with
the SLAC National Accelerator Laboratory.  Additional LSST funding
comes from private donations, grants to universities, and in-kind
support from LSSTC Institutional Members.

\end{document}